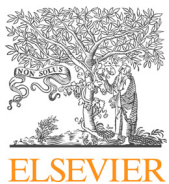

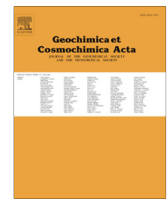

# Surface energetics of wurtzite and sphalerite polymorphs of zinc sulfide and implications for their formation in nature

Tamilarasan Subramani [a], Kristina Lilova [a], Megan Householder [a,b], Shuhao Yang [a], James Lyons [b], Alexandra Navrotsky [a,*]

[a] School of Molecular Sciences and Center for Materials of the Universe, Arizona State University, Tempe, AZ 85287, United States
[b] School of Earth and Space Exploration and Center for Materials of the Universe, Arizona State University, Tempe, AZ 85287, United States

ABSTRACT

Surface energetics of zinc sulfide nanoparticles determines their structure, properties, and occurrence. Using a combination of experimental techniques, we investigated the thermodynamics of the two polymorphs, sphalerite and wurtzite at bulk and nanoscale to understand their occurrence. Calorimetric measurements confirmed that wurtzite has a lower surface energy than sphalerite, which causes a reversal in phase stability at the nanoscale, with wurtzite energetically stable for particle size below 10 nm. Taking these surface energies into account, a simple model of the thermodynamics of the sphalerite - wurtzite transformation as a function of particle size and temperature can explain the occurrence of the zinc sulfide polymorphs in environments as diverse as ore bodies and planetary atmospheres.

© 2022 Elsevier Ltd. All rights reserved.

## 1. Introduction

Zinc sulfide is a common material in diverse geologic environments such as sedimentary exhalative (SEDEX) hydrothermal deposits, Mississippi Valley type deposits (MVT), volcanogenic massive sulfide (VMS) deposits (Nakai et al., 1993; Akbulut et al., 2016; Knorsch et al., 2020). Zinc sulfide nanoparticles have promising applications in semiconductors, sensors, and solar cells (Chen and Lockwood, 2002; He et al., 2007; Fang et al., 2007, 2009; Chen et al., 2008; Löw et al., 2008; Walker et al., 2009) and the interest in their properties has been constantly growing. The dependence of physical properties and stability on particle size and crystal structure affects their manufacturing and functionality. After iron, aluminum, and copper, zinc is the fourth most mined metal in the world and zinc sulfide is the principal mineral ore for its production (Gupta, 2018). The atmospheres of warm and hot exoplanets are likely to contain cloud condensates and photochemical hazes, formed from different chemical species including zinc sulfide (Gao and Benneke, 2018). The nucleation rate of such condensates depends strongly on their surface properties (Gao et al., 2018; Gao and Benneke, 2018) with higher surface energies drastically hindering nucleation. Although nucleation has been modeled extensively by the materials science, geochemistry, and astrophysics communities, the surface energies of many key compounds, including ZnS have not been constrained by experiment.

Zinc sulfide exists as two polymorphs – sphalerite and wurtzite. At atmospheric pressure and room temperature, bulk sphalerite (cubic structure) is slightly more stable than bulk wurtzite (hexagonal structure) (Barin et al., 1977). Although the phase transformation from cubic sphalerite to hexagonal wurtzite is widely accepted to occur at 1020 °C (Allen et al., 1912), there is direct evidence of simultaneous deposition of both phases in ores, showing that natural wurtzite can form at lower temperature (Scott and Barnes, 1972). Moreover, the apparent transition temperature can vary significantly (Scott and Barnes, 1972; Akizuki, 1981), and the thermodynamics of the phase transformation and the phase stabilities of the two polymorphs appear to depend strongly on particle size (Qadri et al., 1999; Zhang et al., 2003b). Transition temperatures as low as 350–400 °C have been reported (Qadri et al., 1999; Zhang et al., 2003a; Huang and Banfield, 2005) when the diameter of the particles is below 10 nm. A possible explanation of these variations is a reversal in phase stability at the nanoscale (i.e. nano wurtzite becomes the more stable phase below some critical particle size or above some critical surface area), similar to reversals seen for the polymorphs of oxides such as $TiO_2$, $ZrO_2$, and $Al_2O_3$ (McHale et al., 1997a, 1997b; Pitcher et al., 2005; Levchenko et al., 2006).

An important factor contributing to the difference in phase stability between the nanophases and their macroscopic analogues could be the presence of various molecules and functional groups (commonly water and/or organics) adsorbed on the surface. The type of adsorbed molecules depends on the synthesis and/or storage conditions. In addition to stabilizing the surface, adsorbed organic molecules can decelerate or completely prevent coarsening (Zhang and Banfield, 2004) and result in retention of small particle size, often below 10 nm (Neouze and Schubert, 2008; Delong et al., 2010; Song et al., 2014; Perla et al., 2020). Relations between the adsorbed molecules on the surface and the particle size and crystal structure are clearly implied by the synthesis methods used to obtain the nanophase ZnS polymorphs. Nano wurtzite only forms in organic solutions/precursors such as ethylene glycol (Biswas and Kar, 2008; La Porta et al., 2013), ethylenediamine (Biswas and Kar, 2008), thiourea (Kole et al., 2014), and thioacetamide (Zhang et al., 2006), while nano sphalerite can be obtained in both aqueous and organic media (Zhang et al., 2006; Biswas and Kar, 2008; Kole et al., 2014; Dengo et al., 2020). According to Zhang et al., solvation (adsorbed methanol) and hydration (adsorbed water) stabilize the sphalerite and wurtzite surfaces differently, resulting in structural transformations at room temperature (Zhang et al., 2003b). The particles formed in methanol have an unidentified structure with poor crystallinity, which changes after methanol desorption. Water adsorption on the same particles increases crystallinity and results in formation of sphalerite. Since organic molecules are present in natural aqueous solutions, similar surface effects may lead to phase selection in geologic and planetary systems.

Here we present an experimental investigation of the effect of particle size on phase formation and thermodynamic stability of the ZnS polymorphs, specifically on the thermodynamics of phase transformation as affected by surface energy. Previously reported values of surface energy are based largely on DFT calculations and range between 0.86 and 1.5 J/m² for sphalerite and between 0.57 and 1.45 J/m² for wurtzite (Nosker et al., 1970; Tauson and Abramovich, 1988; Yoshiyama et al., 1988; Wright et al., 1998; Hamad et al., 2002; Zhang et al., 2003a). Despite the large range of the predicted surface energies, the values suggest that wurtzite may have a slightly smaller surface energy than sphalerite, consistent with its occurrence in nanoparticles (Zhang et al., 2003a) and a possible reversal in stability at small particle size of the two polymorphs. Our calorimetric measurements of the surface energies of the two ZnS polymorphs, combined with previously obtained data (Qadri et al., 1999; Zhang et al., 2003a; Huang and Banfield, 2005), clarify the origin and formation processes of nanophase zinc sulfides in anthropogenic, environmental, geochemical, and planetary settings.

## 2. Materials and methods

### 2.1. Synthesis

Bulk sphalerite (SP1) is a commercial sample with 99.99% purity from Cerac Inc. Nano sphalerite (SP2) was synthesized using a hydrothermal method. In the synthesis, 0.95 mol of zinc nitrate hexahydrate, $Zn(NO_3)_2 \cdot 6H_2O$, and equimolar sodium thiosulfate, $Na_2S_2O_3$, were dissolved in 50 ml of deionized water. The solution was transferred to a 100 ml Teflon container and sealed in a stainless-steel hydrothermal setup. The setup was kept at 115 °C for 18 h. The resulting milky white powder was collected by centrifugation and washed with toluene by stirring at room temperature for several hours. Finally, the powder was washed with ethanol and water alternately several times. Bulk wurtzite (WZ1) was synthesized by heating the bulk sphalerite (SP1) at 1200 °C for 1 h in argon atmosphere and slowly cooled. Nano wurtzite (WZ2) was synthesized by a solvothermal method using ethylene glycol as solvent. $Zn(NO_3)_2 \cdot 6H_2O$ and thiourea were taken in a molar ratio of 1:2 and dissolved in 60 ml of ethylene glycol in a beaker. The mixture was transferred to the hydrothermal setup described above. The setup was placed at different temperatures (110, 150 and 180 °C) to obtain nano wurtzite of different particle sizes.

### 2.2. Characterization

Powder X-ray diffraction (PXRD): PXRD was done on a Bruker D2 benchtop diffractometer equipped with Ni-filtered Cu Kα radiation ($\lambda = 1.542$ Å) operating at an accelerating voltage of 30 kV and emission current of 10 mA. The data were collected in the 2θ range of 10–100° with a 0.018° step size and a 2 s collection time per step. The PXRD pattern was subjected to Rietveld analysis using GSAS-II software (Toby and von Dreele, 2013). VESTA 3 software was used to demonstrate the crystal structures of the different ZnS polymorphs (Momma and Izumi, 2011).

FTIR spectroscopy was done in the region 500–4000 cm$^{-1}$ using a Bruker Vertex 70 spectrometer. Simultaneous thermogravimetry and differential scanning calorimetry (TG-DSC) was performed on a Setaram Labsys Evo instrument. The sample was heated to 600 °C in argon at 10 °C/min and cooled at 20 °C/min. The surface area of the nanophase samples was obtained at 77 K using a 5-point Brunauer–Emmett–Teller (BET) technique on the analysis port of a Micromeritics ASAP 2020 instrument. About 70 mg of the sample was degassed under vacuum at 60 °C for 18 h prior to the N$_2$ adsorption measurement. Adsorption data of N$_2$ in the P/P$_0$ range from 0.05 to 3 were used to fit the BET equation.

Scanning electron microscopy was attempted using a SEM/FIB Focused Ion Beam - Helios 5 UX instrument (Thermo Scientific). Imaging confirmed the occurrence of roughly spherical agglomerates of nanoparticles but the distribution of particle sizes and morphologies could not be quantified because of agglomeration and relatively low resolution. A detailed high resolution transmission electron microscopy study is planned for the future.

### 2.3. High temperature oxidative solution calorimetry

High temperature oxidative solution calorimetry was performed at 1073 K in molten sodium molybdate ($3Na_2O \cdot 4MoO_3$) solvent. Oxygen gas was flushed over the solvent at 90 ml/min and bubbled through it at 5 ml/min. The final product of calorimetry is zinc sulfate dissolved in the molten oxide solvent. The calorimeter was calibrated against the heat of combustion of 5 mg pellets of benzoic acid ($C_7H_6O_2$, Parr Instruments). The calorimetric methodology used to measure the thermodynamic properties on sulfides is described in more detail elsewhere (Abramchuk et al., 2020; Hayun et al., 2020).

### 2.4. Water adsorption calorimetry

Enthalpies of water adsorption were recorded using a Setaram Sensys Evo calorimeter coupled with a Micromeritics ASAP 2020 gas adsorption system used for precise gaseous water dosing and volumetric detection of adsorbed amounts as described previously (Ushakov and Navrotsky, 2005). The sample was placed in one side of a forked silica glass tube and degassed under a static vacuum (<10$^{-6}$ Torr) at 60 °C for 18 h to remove most of the water without coarsening the sample. After the N$_2$ adsorption measurement on the sample and the free space measurement of the tube, the system was re-evacuated. Then, precisely controlled small doses of 0.01 mmol/g of gaseous water were stepwise released into the system at room temperature until P/P₀ reached 0.35. Each dose generated an exothermic calorimetric peak recorded by the calorimeter. The simultaneous record of the amount of adsorbed water and the heat effect provided a high resolution measurement of the differential heat of adsorption as a function of surface coverage.

### 2.5. Coarsening

The coarsening experiments were performed in a Setaram Sensys Evo calorimeter coupled with a Micromeritics ASAP 2010 at 110 and 300 °C in vacuum (<10⁻⁶ Torr) to prevent oxidation.

## 3. Results

### 3.1. Synthesis

The nanophase hydrothermally synthesized ZnS (SP2) formed in the cubic sphalerite structure (Fig. 1a) as expected. Nanophase wurtzite synthesized at 110 °C in ethylene glycol was a poorly crystalline material, not suitable for calorimetry (Fig. 1b) Increasing the synthesis temperature to 150 °C resulted in well crystalized pure phase wurtzite (WZ2) (Fig. 1c). When we increased the synthesis temperature above 150 °C, the triplets in the wurtzite XRD pattern started disappearing, which indicated formation of nano sphalerite as a second phase (Fig. 1d).

### 3.2. Characterization

The BET surface area of SP2 was 60 m²/g, which corresponds to an average particle size of 24 nm and that of WZ2 was 144 m²/g, corresponding to a 10 nm particle size, assuming spherical shape. For comparison, the crystallite size determined from the XRD measurements, using the main characteristic peaks for each phase ((111) for sphalerite and (100), (002), and (101) for wurtzite) and MDI-Jade 9 software is 22 nm for SP2 and 8 nm for WS2. The computation was performed using the Scherrer equation (Scherrer, 1918; Holzwarth and Gibson, 2011) with LaB₆ as a standard for the instrumental line broadening.

The FTIR measurements on nano wurtzite (WZ2) confirmed the presence of ethylene glycol (Fig. S1), while the nano sphalerite (SP2) contained only adsorbed water. The total amount of water in SP2 according to the TG total weight loss was 9.47 wt%, which corresponds to the formula: ZnS·0.56 H₂O. The DSC of WZ2 showed an endothermic peak from 50 to 145 °C and several exothermic peaks above 145 °C. The weight loss (4.2 wt%) in the lower temperature interval was due to the desorption of the physiosorbed water and that in the higher temperature interval (11.7 wt%) to the decomposition of the ethylene glycol. The formula calculated from these data is ZnS·0.23 H₂O·0.21(CH₂OH)₂.

### 3.3. Water adsorption

The enthalpy of water adsorption was directly measured as a function of the amount of water adsorbed for both polymorphs. The adsorption enthalpies for individual doses (differential enthalpies of adsorption) became less exothermic with successive water doses and eventually reached the enthalpy of bulk water condensation (−44.0 kJ/mol), corresponding to a change from chemisorption to physisorption, see Fig. 2a. The chemisorbed water is approximately-one water monolayer if one compares the coverage to values calculated by DFT (Dengo et al., 2020, Sun et al. (2013) and Li et al., 2019, Simpson et al., 2011). For nano sphalerite, chemisorbed water corresponds to coverage < 20 $H_2O$/ $nm^2$. The remaining water is treated as physically adsorbed water. The integral enthalpy, which is the sum of the differential enthalpies of adsorption divided by the total amount of adsorbed water up to the coverage where the differential enthalpy reached −44 kJ/ mol, gives the total adsorption enthalpy for chemisorbed water as −65 kJ/mol. The adsorption energy calculated by DFT for an ideal monolayer on the (110) nano sphalerite is between −61.5 and −80.7 kJ/mol (Dengo et al., 2020). Sun et al. (2013) and Li et al., (2019) present more consistent values of −74.7 and −70.8 kJ/mol, respectively, while Simpson et al. calculated an energy of −77 kJ/ mol (Simpson et al., 2011). Overall, we consider the agreement between theory and experiment to be satisfactory. The water adsorption enthalpy of WZ2 was measured (Fig. 2a) following the same procedure, resulting in an integral value of −55 kJ/mol, which is less exothermic than the experimentally measured and calculated energies for nano sphalerite. The less negative differential heats of adsorption of nano wurtzite are consistent with it having a lower surface energy and less tightly bound water than nano sphalerite.

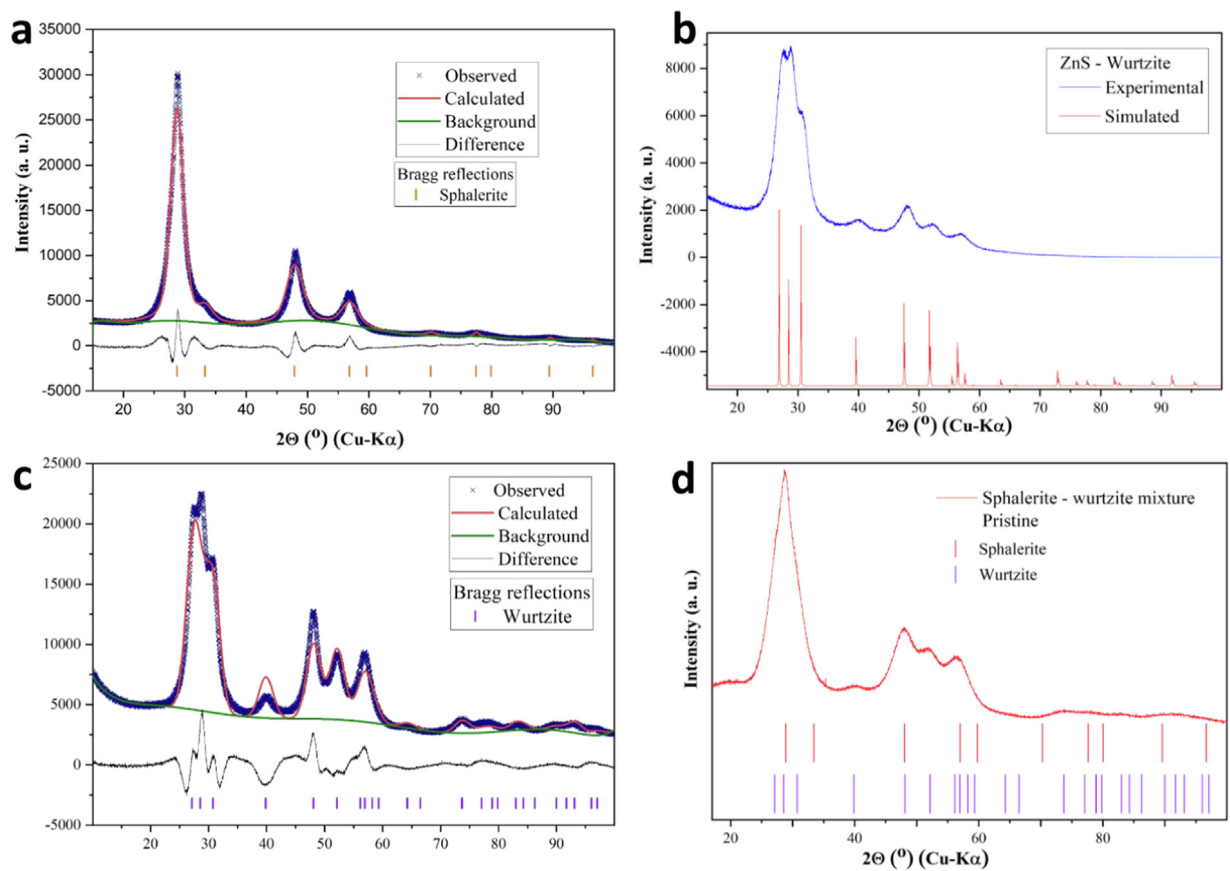

**Fig. 1.** PXRD pattern of nano sphalerite and nano wurtzite samples. (a) Rietveld refinement profile of nano-sphalerite (SP2) made in water at 110 °C for 18 h. Reliability factors: R = 7.33%, wR = 9.59% and GOF = 7.14. (b) PXRD pattern of wurtzite ZnS made at 110 °C/18hrs with the lowest particle size (<10 nm). The ratio of intensity of first two peaks of triplet is almost equal indicating the phase is wurtzite. (c) Rietveld refinement profile of nano-wurtzite (WZ2) made in ethylene glycol at 150 °C for 18 h. Reliability factors: R = 7.51%, wR = 9.09% and GOF = 5.09. (d) PXRD of ZnS sample made at 180 °C/18hrs showing mixture of sphalerite and wurtzite.

The differential heats of adsorption were used to calculate the enthalpies of hydration of sphalerite and wurtzite, namely for the reaction:

$$ZnS \text{ (anhydrous)} + xH_2O \text{ (liquid)} \rightarrow ZnS \cdot xH_2O(1)$$

As a function of the amount of adsorbed water (Fig. 2b). This was done by taking the integral of the adsorption enthalpies from the first dose to each intermediate water content and correcting for the enthalpy of condensation of water vapor. These values level off for higher water contents once the heat of adsorption becomes equal to the heat of condensation of water vapor. Thus water addition in the physisorption regime, does not further stabilize the wurtzite or sphalerite nanoparticles. The largest energetic stabilization due to hydration is about −1 kJ/mol for wurtzite and −4 kJ/mol for sphalerite. These effects are very small in magnitude, so entropy terms may also play a role in stabilization, but these cannot be evaluated at present. The enthalpy of hydration indicates how the absorbed water increases the stability of the nanophase by decreasing the excess energy of its surface. The less stabilizing effect of water on nano wurtzite than on nano sphalerite could explain why nano wurtzite does not form readily in aqueous solutions. Further measurements of the enthalpy of solvation as a function of the amount of organics adsorbed on both polymorphs could provide more insight into the gas–solid interactions on their surfaces and the effect of those interactions on the resulting structure and particle size.

### 3.4. Coarsening

Since the particle size effect on the sphalerite to wurtzite transformation has been studied (Qadri et al., 1999; Zhang et al., 2003a; Huang and Banfield, 2005), we focused on the reverse phase transition (wurtzite to sphalerite) through a series of annealing (coarsening) experiments. The average particle size of each phase and the ratio of the phases were estimated from the XRD measurements. When the nano wurtzite (WZ2) sample with average particle size of 10 nm was annealed at 110 °C for 15 h, a decrease in the intensity of the XRD triplets characteristic of wurtzite was observed, indicating a small amount (<10%) of newly formed sphalerite (Fig. 3a). Increasing the temperature to 300 °C resulted in 50% sphalerite with particle size of 20 nm (Fig. 3b). The mixed phase sphalerite – wurtzite sample, prepared initially at 180 °C, was then annealed at 300 °C for 6 h to assess the coarsening behavior when both phases were present (Fig. 3c). The ratio and the average particle size of the nano sphalerite and nano wurtzite changed from 50:50 and 18 and 8 nm to 80:20 and 25 and 10 nm, respectively.

### 3.5. High temperature oxidative solution calorimetry, enthalpies of formation, and surface enthalpies

The thermodynamic stability of all four samples was measured by oxidative solution calorimetry. The heat effect obtained for each compound was used to calculate their enthalpies of formation from elements and the surface energies of the nano ZnS (Table 1). The solution enthalpies of the nanophase samples were corrected for the effect of the molecules adsorbed on the surface (SI). The thermodynamic cycles used to correct the drop solution enthalpies and the enthalpies of formation are shown in SI.

The enthalpies of formation of the bulk phases are consistent with the literature values (Robie and Hemingway, 1995) within experimental error. The transition enthalpy between bulk spha-lerite and bulk wurtzite, calculated as the difference between their solution enthalpies, is $3.61 \pm 1.98$ kJ/mol, which is in good agreement with the literature value of $2.5 \pm 1.5$ kJ/mol (Gardner and Pang, 1988).

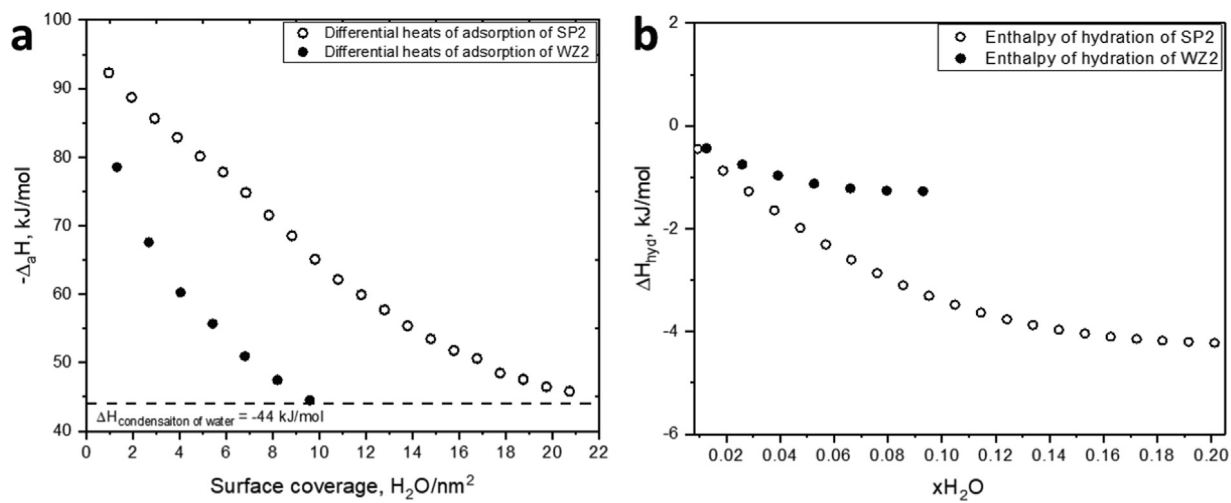

**Fig. 2.** (a) Measured differential enthalpy of adsorption of water on sphalerite (SP2) and wurtzite (WZ2) as a function of coverage ($H_2O$ /$nm^2$), (b) The enthalpy of hydration (liquid water reference state) of nanophase sphalerite (SP2) and wurtzite (WZ2) as a function of the amount of water.

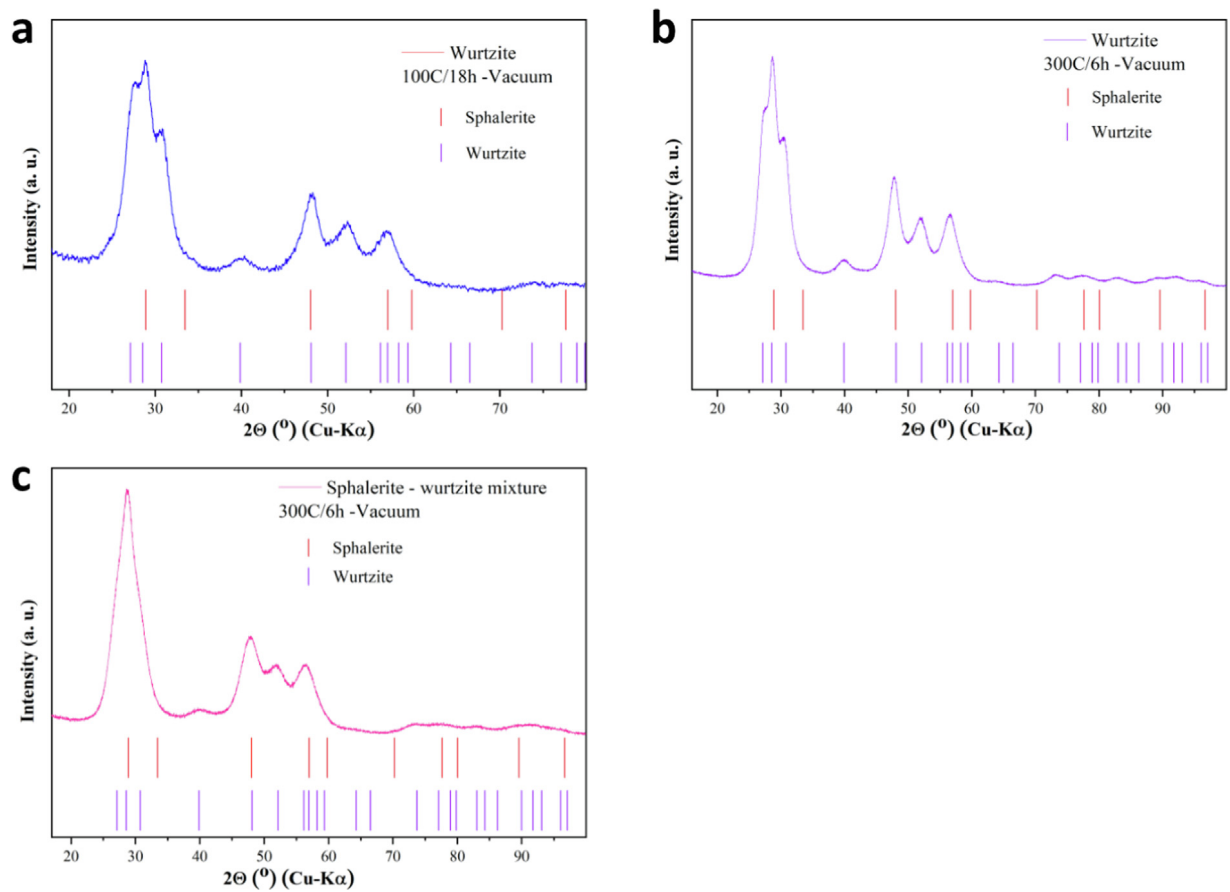

**Fig. 3.** Particle size effect on the reverse phase transition. (a) PXRD of wurtzite ZnS sample after annealing at 110 °C/18 h under vacuum. The intensity of the second peak of the triplet which is the prominent signature of sphalerite increased, indicating the transformation of some wurtzite particles into sphalerite. (b) PXRD of wurtzite ZnS sample after annealing at 300 °C/6 h under vacuum. The intensity of the second peak of the triplet increased further, indicating transformation of more wurtzite particles into sphalerite. (c) PXRD of sphalerite and wurtzite mixture after annealing at 300 °C/6 h under vacuum. The intensities of peaks corresponding to sphalerite increased.

**Table 1**
Measured enthalpies of the ZnS samples.

| Sample | $\Delta H_{ds}$ (kJ/mol)[a] | $\Delta H_{f,el}$ (kJ/mol)[b] | $\Delta H_{f,el}$ (kJ/mol)[c] |
|---|---|---|---|
| Bulk sphalerite (SP1) | $-767.61 \pm 1.36$ (10)[d] | $-204.03 \pm 3.57$ | $-204.1 \pm 1.5$ |
| Nano sphalerite (SP2) | $-774.78 \pm 1.27$ (8) | $-196.86 \pm 3.53$ | |
| Bulk wurtzite (WZ1) | $-771.22 \pm 1.44$ (8) | $-200.42 \pm 3.60$ | $-203.8 \pm 1.5$ |
| Nano wurtzite (WZ2) | $-785.04 \pm 4.29$ (8) | $-185.60 \pm 5.41$ | |

[a] Drop solution enthalpies of bulk and nano sphalerite and wurtzite.
[b] Enthalpies of formation of bulk ZnS polymorphs from elements (this work).
[c] Enthalpies of formation of bulk ZnS polymorphs from elements (Robie and Hemingway, 1995).
[d] Uncertainty is two standard deviations of the mean, value in ( ) is number of experiments performed.

The surface enthalpy of the hydrated/solvated surface is $1.25 \pm 0.21$ J/m² for sphalerite and $0.99 \pm 0.32$ J/m² for wurtzite. Both experimental values are within the computationally predicted range and indicate stabilization of nano wurtzite relative to nano sphalerite (Nosker et al., 1970; Tauson and Abramovich, 1988; Yoshiyama et al., 1988; Wright et al., 1998; Hamad et al., 2002). Using the same approach, the surface enthalpy of anhydrous nano sphalerite is calculated as $1.76 \pm 0.32$ J/m² (SI). If we assume a similar difference between the surface enthalpies of the solvated and unsolvated surface of wurtzite, the surface enthalpy of unsolvated nano wurtzite can be estimated as $1.50 \pm 0.50$ J/m².

## 4. Discussion

### 4.1. Stability landscape and enthalpy crossover

Thermodynamic phase stability is governed by the Gibbs free energy. Previous studies have shown that the entropies of several nanophase polymorphs are similar to those of bulk materials (Boerio-Goates et al., 2006); thus the surface enthalpies are the leading terms in calculating the nanoscale stability landscape. The calculation similar to that done for other nanophases (McHale et al., 1997a, 1997b; Pitcher et al., 2005; Levchenko et al., 2006), is shown in the SI. The results, showing enthalpy relative to bulk sphalerite as a function of surface area, are shown in Fig. 4. A definite crossover in enthalpy is seen at a particle size of 9.8 nm.

This calculated crossover at a particle size near 10 nm is consistent with previous experimental data, which show that nano sphalerite with a diameter $\leq 9$ nm will start converting to wurtzite at 350 °C and will transform almost completely at 500 °C. Similar to nanophase oxides (McHale et al., 1997a, 1997b; Pitcher et al., 2005; Levchenko et al., 2006), the general trend of energy crossovers supports the observation that polymorphs with lower surface energy (enthalpy) become stable at the nanoscale. The high temperature wurtzite ZnS polymorph has lower surface enthalpy and less exothermic heat of water adsorption, while the thermodynamically stable low temperature sphalerite polymorph has higher surface enthalpy and more exothermic water adsorption enthalpy. When a small particle of nano sphalerite is annealed and most of the water adsorbed on the surface is eliminated without significant coarsening (Zhang et al., 2003a), the zinc sulfide transforms from cubic (sphalerite) to hexagonal (wurtzite), which is the stable polymorph at that particle size. It has been established that nano wurtzite cannot be synthesized in aqueous solution, but only in organic media/precursors (Zhang et al., 2006; Biswas and Kar, 2008, La Porta et al., 2013, Kole et al., 2014) and the formation process can occur with several different organics. This lack of specificity suggests that the organic does not need a lock-and-key fit to some particular surface site. Rather, the large organic molecules attach to the surface and block contact between adjacent particles, and prevent diffusion and coarsening, thus keeping the particle size within the stability range of nano wurtzite. Upon heating above 100 °C, the largest wurtzite particles (close to 10 nm diameter) start losing the surface molecules and converting to sphalerite, which is stable at larger particle size. All our annealing experiments have shown that the remaining wurtzite retains its initial particle size (between 8 and 10 nm), while the newly formed sphalerite is around 20–25 nm depending on the temperature. It can be concluded that at the nano level the sphalerite to wurtzite and wurtzite to sphalerite transformations occur at significantly lower temperatures than for the bulk compounds and, importantly, the processes are governed mainly by thermodynamics and often are reversible.

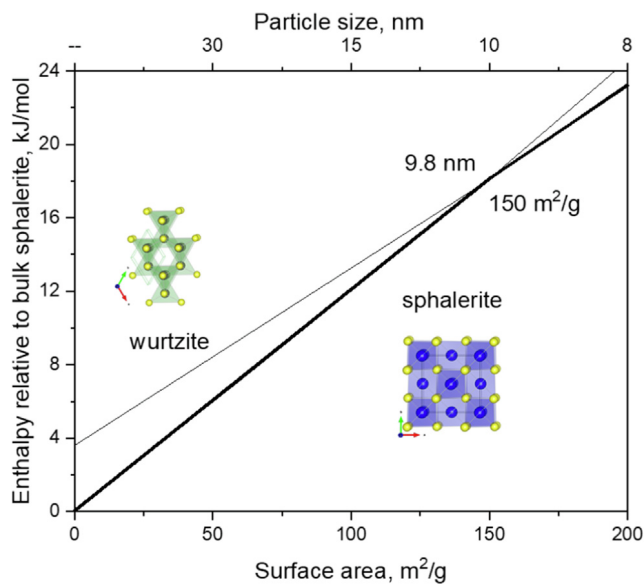

**Fig. 4.** Phase stability crossover calculated using our experimental thermodynamic data for nanocrystalline ZnS. The darker line segments indicate the energetically stable phases.

An example of wurtzite nucleation based on surface energy and the stability crossover at the nanoscale is seen in III-V nanowire semiconductors such as InAs and GaAs. Pure wurtzite structure is observed with a diameter up to 10 nm (Shtrikman et al., 2009). According to Glas et al. (2007), the nanowires initially start growing as sphalerite with triangular bases and tilted lateral facets. Subsequent growth results in a transition to the sphalerite structure.

### 4.2. Geochemical implications of the surface energetics for the formation of zinc sulfide ore deposits

The phase transformation from bulk sphalerite to bulk wurtzite occurs at 1020 °C (Barin et al., 1977), which is also approximately the temperature at which we obtained bulk wurtzite by annealing commercial bulk sphalerite. The process is claimed to be thermodynamically reversible (Scott and Barnes, 1972) but our sample remained pure wurtzite even after being slowly cooled, which indicates that the reverse (wurtzite to sphalerite) transition is kinetically hindered in the absence of an aqueous solution. Most of the natural deposits of zinc sulfide contain both polymorphs in a multiphase assemblage in equilibrium with other minerals, which is evidence of natural wurtzite formation at temperatures below 1020 °C (Scott and Barnes, 1972; Akizuki, 1981). Since particle size affects the transition temperature as discussed above, the explanation for this variability in transition temperature could be that both polymorphs are formed initially as nanoparticles and that the transformation between them is determined by the surface energetics. Since precipitation of sulfides in nature is often a rapid process driven by changes in concentrations and pH, nanoparticles are indeed likely to form initially. Fig. 5 shows the possible sequence of formation of the natural ZnS ores from nanoparticles.

Earlier studies assumed that bulk wurtzite forms as a metastable phase in low temperature hydrothermal solutions in shallow subaerial or submarine conditions as a result of the rapid mixing of hydrothermal fluid with cold seawater (Sunagawa, 1984; Ikehata et al., 2015) and then transforms to sphalerite with time depending on the temperature and pressure (Kojima and Ohmoto, 1991). Another hypothesis is that natural wurtzite forms in low temperature hydrothermal conditions because it is more stable than sphalerite at low sulfur fugacity (Scott and Barnes, 1972). (Gartman et al., 2019) demonstrated recently that both zinc sulfide polymorphs form at the interface of hydrothermal fluids and ocean water initially as nanoparticles and appear to remain nanosized even as aggregates. Experimental studies have shown that the coarsening of nano ZnS is slow even at 225 °C (Gilbert et al., 2003) and that colloidal ZnS (perhaps being poorly crystalline wurtzite) and sphalerite in sulfidic solutions at room temperature do not equilibrate for months (Daskalakis and Helz, 1993). Another factor stabilizing wurtzite at the nanoscale could be the increased organic content of the aqueous solution arising from microbial processes and from the rapid cooling of hydrothermal fluids upon emission into seawater (Breier et al., 2012). Such organics could interact with the surface of the nanoparticles, stabilizing the wurtzite phase and suppressing coarsening until there is a change in environmental conditions. Additionally, some of the smallest ZnS nanoparticles (<5 nm) can be biogenic in origin, for example produced by sulfate-reducing bacteria (Moreau et al., 2004). It has been observed previously that a mixture of cubic and hexagonal ZnS crystals usually forms at lower temperatures (Smith, 1955), which supports the hypothesis that they are both nanosized with wurtzite being stable below 10 nm. However, the previously suggested rarity of natural wurtzite (Leach et al., 2010) may be an artefact of inaccurate reporting since wurtzite is easily overlooked and misidentified as sphalerite in mixed fine-grained deposits (Leach et al., 2010; Pring et al., 2020).

Since we have shown thermodynamically that the coarsening of an initially precipitated nano wurtzite produced abiogenically or within a biofilm (Labrenz et al., 2000; Moreau et al., 2004) would not result in the formation of bulk wurtzite, we present a somewhat different possible sequence of the formation process (Fig. 5).

Upon increasing to a particle size around 10 nm, nano wurtzite is thermodynamically favored to transform to nano sphalerite. If nano sphalerite with very small particle size (3–9 nm) is initially formed, it is thermodynamically favored to transform to nano wurtzite as shown by the current work and previous works (Qadri et al., 1999; Zhang et al., 2003a; Huang and Banfield, 2005). Thus, the sphalerite - wurtzite transition at the nanoscale is reversible and can go in either direction depending on the initial phase formed and particle size, the competition between rates of coarsening and of phase transformation. The slow coarsening and transformations of the precipitated nanosized zinc sulfides over geologic time are likely to result in predominantly bulk sphalerite with traces of wurtzite, as found in sedimentary exhalative (SEDEX) hydrothermal deposits (Knorsch et al., 2020), Mississippi-Valley type deposits (MVT) (Viets et al., 1992; Nakai et al., 1993), volcanogenic massive sulfide (VMS) deposits (Akbulut et al., 2016), and in mixed bulk cubic – hexagonal zinc sulfide minerals (Leach et al., 2010). The sequence in Fig. 5 is especially relevant to the formation of the ZnS polymorphs when hydrothermal fluids meet ocean water and thus for volcanogenic massive sulfide (VMS) deposits. Since coarsening of nanophase wurtzite produces sphalerite, the bulk wurtzite found in natural ore deposits is almost certainly a result of subsequent transformation of bulk sphalerite to wurtzite at higher temperature. It should be noted that the transition temperature of pure sphalerite to wurtzite – 1020 °C (Scott and Barnes, 1972) is above the formation temperature of most sulfide ores (Candela, 2003). Those natural ores are never chemically pure phases and usually contain impurities like Fe, Mn, Cd, and Cu, which lower the sphalerite – wurtzite transformation temperature to 350–900 °C (Maurel, 1978; Tauson, 1978, Tauson and Chernyshev, 1978; Shima et al., 1982; Osadchii, 1991; Kubo et al., 1992). The Fe–Zn–S system is considered one of the best geothermometers for natural sulfide ores and the phase relations among sphalerite, pyrite, pyrrhotite, and wurtzite have been studied extensively (Kullerud, 1953; Fleet, 2006). Sphalerite containing 56 mol.% Fe substituting for Zn transforms to wurtzite around 850 °C (Kullerud, 1953; Barton and Toulmin, 1966). 26 mol.% of Mn in the Mn - Zn - S solid solution can decrease the transition temperature to 350 °C (Tauson and Chernyshev, 1978). (Maurel, 1978) investigated the effect of Cd content on the occurrence of the ZnS polymorphs at 650 and 800 °C and found that sphalerite forms below 12.2 mol.% Cd, both phases coexist between 12.2 and 14.3 mol.%, and wurtzite is the only phase formed above 14.3 mol.% Cd. However, it is not clear whether these amounts represent equilibrium solubilities and what other phases may buffer the extent of substitution. Further thermodynamic studies of the mixing properties of the bulk and nanoscale solid solutions of ZnS with FeS, MnS and CdS are needed to constrain the thermodynamic properties and provide insight into the stability of different structures in these solid solutions.

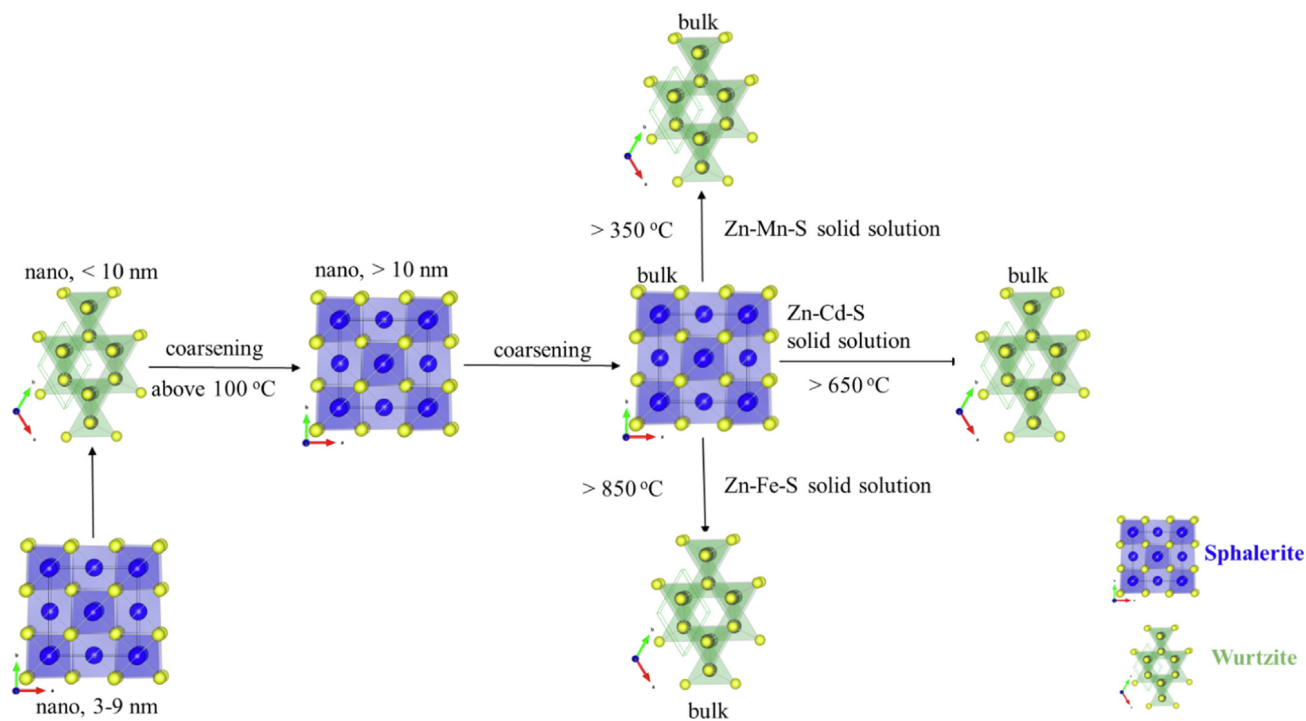

**Fig. 5.** Possible formation of polymorphs from nanosized sphalerite and wurtzite in the presence of impurities.

### 4.3. Implications beyond Earth

Until this study it was assumed that sphalerite was the dominant zinc sulfide polymorph in hot exoplanet (Kopparapu et al., 2018; Gao and Benneke, 2018) but our work suggests that the formation process might follow a path as shown in Fig. 5. ZnS precipitating rapidly from a gas phase in a planetary atmosphere (or elsewhere) is likely to form initially as a poorly crystalline or amorphous nanophase, and as the cloud condensation nucleus grows, that phase transforms to fine nanocrystalline wurtzite, followed by nano and bulk sphalerite due to coarsening, and then the condensate would transform to bulk wurtzite if the temperature were high enough. Due to the coarsening kinetics, the origin of the particles, and the environmental conditions, the mechanism of the formation process is possibly more complex than our simplified schematic suggests and the final product may not be a single phase polymorph, but a mixture of wurtzite and sphalerite with different particle sizes and possibly compositions. Despite this possible complexity, the experimentally measured surface energies of sphalerite and wurtzite, combined with nucleation rate and particle size distribution calculations, can be used to model the composition and phases present in these exotic atmospheres using microphysics and radiative transfer codes. Using such models, the aerosol absorption and scattering in exoplanet transmission and emission spectral models can be more accurately evaluated (Brogi and Line, 2019). With the James Webb Space Telescope's mirrors now aligned and producing new spectral data from exoplanets never seen before, the need for experimentally measured surface energetics has never been more critical for the astrophysical community.

## 5. Conclusions

In summary our study explores the connection between the surface energetics of the two ZnS polymorphs, sphalerite and wurtzite, and their existence and possible formation process in ore deposits and in exoplanet atmospheres. The reversal of phase stability at the nanoscale can explain and predict which structure will be prevalent under any given environmental conditions.

## Data availability

Data will be made available on request.

## Declaration of Competing Interest

The authors declare that they have no known competing financial interests or personal relationships that could have appeared to influence the work reported in this paper.

## Acknowledgments

This work was supported by the U.S. Department of Energy, Office of Basic Energy Sciences, grant DE-FG02-97ER14749.

## Appendix A. Supplementary material

Supplementary material to this article can be found online at https://doi.org/10.1016/j.gca.2022.11.003.

## References

Abramchuk, M., Lilova, K., Subramani, T., Yoo, R., Navrotsky, A., 2020. Development of high-temperature oxide melt solution calorimetry for p-block element containing materials. J. Mater. Res. 35, 2239–2246.

Akbulut, M., Oyman, T., Çiçek, M., Selby, D., Özgenç, İ., Tokçaer, M., 2016. Petrography, mineral chemistry, fluid inclusion microthermometry and Re–Os geochronology of the Küre volcanogenic massive sulfide deposit (Central Pontides, Northern Turkey). Ore Geol. Rev 76, 1–18.

Akizuki, M., 1981. Investigation of phase transition of natural ZnS minerals by high resolution electron microscopy. Am. Miner. 66, 1006–1012.

Allen, E.T., Crenshaw, J.L., Merwin, H.E., 1912. The sulphides of zinc, cadmium, and mercury; their crystalline forms and genetic conditions. Am. J. Sci. s4, 34, 341–396.

Barin, I., Knacke, O., Kubaschewski, O., 1977. Thermochemical properties of inorganic substances. Springer, Berlin Heidelberg.

Barton, P.B., Toulmin, P., 1966. Phase relations involving sphalerite in the Fe-Zn-S system. Econ. Geol. 61, 815–849.

Biswas, S., Kar, S., 2008. Fabrication of ZnS nanoparticles and nanorods with cubic and hexagonal crystal structures: a simple solvothermal approach. Nanotechnology 19, 045710, pp. 1–11.

Boerio-Goates, J., Li, G., Li, L., Walker, T.F., Parry, T., Woodfield, B.F., 2006. Surface water and the origin of the positive excess specific heat for 7 nm rutile and anatase nanoparticles. Nano Lett. 6, 750–754.

Breier, J.A., Toner, B.M., Fakra, S.C., Marcus, M.A., White, S.N., Thurnherr, A.M., German, C., 2012. Sulfur, sulfides, oxides and organic matter aggregated in submarine hydrothermal plumes at 9°50′N East Pacific Rise. Geochim. Cosmochim. Acta 88, 216–236.

Brogi, M., Line, M.R., 2019. Retrieving temperatures and abundances of exoplanet atmospheres with high-resolution cross-correlation spectroscopy. Astron. J. 157, 114, pp. 1–17.

Candela, P.A., 2003. 3.12 – Ores in the Earth's crust. In: Holland, H.D., Turekian, K.K. (Eds.), Treatise on Geochemistry. Pergamon, Oxford, pp. 411–431.

Chen, Z.-G., Zou, J., Liu, G., Lu, H.F., Li, F., Lu, G.Q., Cheng, H.M., 2008. Silicon-induced oriented ZnS nanobelts for hydrogen sensitivity. Nanotechnology 19, 055710, pp. 1–5.

Chen, R., Lockwood, D.J., 2002. Developments in luminescence and display materials over the last 100 years as reflected in electrochemical society publications. J. Electrochem. Soc. 149, S69–S77.

Daskalakis, K.D., Helz, G.R., 1993. The solubility of sphalerite (ZnS) in sulfidic solutions at 25°C and 1 atm pressure. Geochim. Cosmochim. Acta 57, 4923–4931.

Delong, R.K., Reynolds, C.M., Malcolm, Y., Schaeffer, A., Severs, T., Wanekaya, A., 2010. Functionalized gold nanoparticles for the binding, stabilization, and delivery of therapeutic DNA, RNA, and other biological macromolecules. Nanotechnol. Sci. Appl. 3, 53–63.

Dengo, N., Vittadini, A., Natile, M.M., Gross, S., 2020. In-depth study of ZnS nanoparticle surface properties with a combined experimental and theoretical approach. J. Phys. Chem. C 124, 7777–7789.

Fang, X.S., Bando, Y., Shen, G.Z., Ye, C.H., Gautam, U.K., Costa, P.M.F.J., Zhi, C.Y., Tang, C.C., Golberg, D., 2007. Ultrafine ZnS nanobelts as field emitters. Adv. Mater. 19, 2593–2596.

Fang, X., Bando, Y., Liao, M., Gautam, U.K., Zhi, C., Dierre, B., Liu, B., Zhai, T., Sekiguchi, T., Koide, Y., Golberg, D., 2009. Single-crystalline ZnS nanobelts as ultraviolet-light sensors. Adv. Mater. 21, 2034–2039.

Fleet, M.E., 2006. Phase equilibria at high temperatures. Rev. Mineral. Geochem. 61, 365–419.

Gao, P., Benneke, B., 2018. Microphysics of KCl and ZnS clouds on GJ 1214 b. Astrophys. J 863, 165, pp. 1–23.

Gao, P., Marley, M.S., Ackerman, A.S., 2018. Sedimentation efficiency of condensation clouds in substellar atmospheres. Astrophys. J 855, 86, pp. 1–15.

Gardner, P.J., Pang, P., 1988. Thermodynamics of the zinc sulphide transformation, sphalerite → wurtzite, by modified entrainment. J. Chem. Soc., Faraday Trans. 1 84, 1879–1887.

Gartman, A., Findlay, A.J., Hannington, M., Garbe-Schönberg, D., Jamieson, J.W., Kwasnitschka, T., 2019. The role of nanoparticles in mediating element deposition and transport at hydrothermal vents. Geochim. Cosmochim. Acta 261, 113–131.

Gilbert, B., Zhang, H., Huang, F., Finnegan, M.P., Waychunas, G.A., Banfield, J.F., 2003. Special phase transformation and crystal growth pathways observed in nanoparticles. Geochem. Trans. 4, 20–27.

Glas, F., Harmand, J.C., Gilles, P., 2007. Why does wurtzite form in nanowires of III-V zinc blende semiconductors? Phys. Rev. Lett. 99, 146101, pp. 1–4.

Gupta, A., 2018. Materials: Abundance, purification, and the energy cost associated with the manufacture of Si, CdTe, and CIGS PV. In: Letcher, T.M., Fthenakis, V.M. (Eds.), A comprehensive guide to solar energy systems. Academic Press, Cambridge, MA, USA, pp. 445–467.

Hamad, S., Cristol, S., Catlow, C.R.A., 2002. Surface structures and crystal morphology of ZnS: Computational study. J. Phys. Chem. B 106, 11002–11008.

Hayun, S., Lilova, K., Salhov, S., Navrotsky, A., 2020. Enthalpies of formation of high entropy and multicomponent alloys using oxide melt solution calorimetry. Intermetallics 125, 106897, pp. 1–7.

He, J.H., Zhang, Y.Y., Liu, J., Moore, D., Bao, G., Wang, Z.L., 2007. ZnS/silica nanocable field effect transistors as biological and chemical nanosensors. J. Phys. Chem. C 111, 12152–12156.

Holzwarth, U., Gibson, N., 2011. The Scherrer equation versus the 'Debye-Scherrer equation'. Nature Nanotech. 6, 534.

Huang, F., Banfield, J.F., 2005. Size-dependent phase transformation kinetics in nanocrystalline ZnS. J. Am. Chem. Soc. 127, 4523–4529.

Ikehata, K., Suzuki, R., Shimada, K., Ishibashi, J., Urabe, T., 2015. Mineralogical and geochemical characteristics of hydrothermal minerals collected from hydrothermal vent fields in the Southern Mariana spreading center. In: Ishibashi, J., Okino, K., Sunamura, M. (Eds.), Subseafloor Biosphere Linked to Hydrothermal Systems: TAIGA Concept. Springer Japan, Tokyo, pp. 275–287.

Knorsch, M., Nadoll, P., Klemd, R., 2020. Trace elements and textures of hydrothermal sphalerite and pyrite in Upper Permian (Zechstein) carbonates of the North German Basin. J. Geochem. Explor. 209, 106416.

Kojima, S., Ohmoto, H., 1991. Hydrothermal synthesis of wurtzite and sphalerite at T=350°-250°C. Min. Geol. 41, 313–327.

Kole, A.K., Sekhar, T.C., Kumbhakar, P., 2014. Morphology controlled synthesis of wurtzite ZnS nanostructures through simple hydrothermal method and observation of white light emission from ZnO obtained by annealing the synthesized ZnS nanostructures. J. Mater. Chem. C 2, 4338–4346.

Kopparapu, R.K., Hébrard, E., Belikov, R., Batalha, N.M., Mulders, G.D., Stark, C., Teal, D., Domagal-Goldman, S., Mandell, A., 2018. Exoplanet classification and yield estimates for direct imaging missions. Astrophys. J. 856, 122, pp. 1–12.

Kubo, T., Nakato, T., Uchida, E., 1992. An experimental study on partitioning of Zn, Fe, Mn and Cd between sphalerite and aqueous chloride solution. Min. Geol. 42, 301–309.

Kullerud, G., 1953. The FeS-ZnS system: A geological thermometer. Nor. Geol. Tidsskr. 32, 62–147.

La Porta, F.A., Ferrer, M.M., de Santana, Y.V.B., Raubach, C.W., Longo, V.M., Sambrano, J.R., Longo, E., Andrés, J., Li, M.S., Varela, J.A., 2013. Synthesis of wurtzite ZnS nanoparticles using the microwave assisted solvothermal method. J. Alloys Compd. 556, 153–159.

Labrenz, M., Druschel, G.K., Thomsen-Ebert, T., Gilbert, B., Welch, S.A., Kemner, K.M., Logan, G.A., Summons, R.E., Stasio, G.D., Bond, P.L., Lai, B., Kelly, S.D., Banfield, J. F., 2000. Formation of sphalerite (ZnS) deposits in natural biofilms of sulfate-reducing bacteria. Science 290, 1744–1747.

Leach, D.L., Taylor, R.D., Fey, D.L., Diehl, S.F., Saltus, R.W., 2010. A deposit model for Mississippi Valley-Type lead-zinc ores: Chapter A in mineral deposit models for resource assessment. Sci. Investig. Rep. 2010-5070-A.

Levchenko, A.A., Li, G., Boerio-Goates, J., Woodfield, B.F., Navrotsky, A., 2006. TiO₂ stability landscape: polymorphism, surface energy, and bound water energetics. Chem. Mater. 18, 6324–6332.

Li, Y., Chen, J., Chen, Y., Zhu, Y., Liu, Y., 2019. DFT simulation on interaction of H₂O molecules with ZnS and Cu-activated surfaces. J. Phys. Chem. C 123, 3048–3057.

Löw, P., le Pioufle, B., Kim, B., Bergaud, C., 2008. Assembly of CdSe/ZnS nanocrystals on microwires and nanowires for temperature sensing. Sens. Actuators B 130, 175–180.

Maurel, C., 1978. Stabilité de la blende dans le système Zn-Cd-S. Bull. Minéral. 101, 406–411.

McHale, J.M., Auroux, A., Perrotta, A.J., Navrotsky, A., 1997a. Surface energies and thermodynamic phase stability in nanocrystalline aluminas. Science 277, 788–791.

McHale, J.M., Navrotsky, A., Perrotta, A.J., 1997b. Effects of increased surface area and chemisorbed H$_2$O on the relative stability of nanocrystalline γ-Al$_2$O$_3$ and α-Al$_2$O$_3$. J. Phys. Chem. B 101, 603–613.

Momma, K., Izumi, F., 2011. VESTA 3 for three-dimensional visualization of crystal, volumetric and morphology data. J. Appl. Crystallogr. 44, 1272–1276.

Moreau, J.W., Webb, R.I., Banfield, J.F., 2004. Ultrastructure, aggregation-state, and crystal growth of biogenic nanocrystalline sphalerite and wurtzite. Am. Miner. 89, 950–960.

Nakai, S., Halliday, A.N., Kesler, S.E., Jones, H.D., Kyle, J.R., Lane, T.E., 1993. Rb-Sr dating of sphalerites from Mississippi Valley-type (MVT) ore deposits. Geochim. Cosmochim. Acta 57, 417–427.

Neouze, M.-A., Schubert, U., 2008. Surface modification and functionalization of metal and metal oxide nanoparticles by organic ligands. Monatsh. Chem. 139, 183–195.

Nosker, R.W., Mark, P., Levine, J.D., 1970. Polar surfaces of wurtzite and zincblende lattices. Surf. Sci. 19, 291–317.

Osadchii, E., 1991. The kesterite-cernyite and sphalerite-greenockite solid solutions in the system Cu$_2$SnS$_3$-ZnS-CdS at 400°C and 101.3 MPa. Neu. Jb. Mineral. Monatsh, 457–463.

Perla, V.K., Ghosh, S.K., Pal, T., Mallick, K., 2020. Organic molecule stabilized bismuth sulfide nanoparticles: A potential system for bistable resistive memory application. Physica E Low Dimens. Syst. Nanostruct. 116, 113787, pp. 1–6.

Pitcher, M.W., Ushakov, S.V., Navrotsky, A., Woodfield, B.F., Li, G., Boerio-Goates, J., Tissue, B.M., 2005. Energy crossovers in nanocrystalline zirconia. J. Am. Ceram. Soc. 88, 160–167.

Pring, A., Wade, B., McFadden, A., Lenehan, C.E., Cook, N.J., 2020. Coupled substitutions of minor and trace elements in co-existing sphalerite and wurtzite. Minerals 10, 147, pp. 1–14.

Qadri, S.B., Skelton, E.F., Hsu, D., Dinsmore, A.D., Yang, J., Gray, H.F., Ratna, B.R., 1999. Size-induced transition-temperature reduction in nanoparticles of ZnS. Phys. Rev. B 60, 9191–9193.

Robie, R., Hemingway, B.S., 1995. Thermodynamic properties of minerals and related substances at 298.15K and 1 Bar, USGS Publications Warehouse, U.S. Geological Survey, Information Services.

Scherrer, P., 1918. Göttinger Nachrichten. Math. Phys. 2, 98–100.

Scott, S.D., Barnes, H.L., 1972. Sphalerite-wurtzite equilibria and stoichiometry. Geochim. Cosmochim. Acta 36, 1275–1295.

Shima, H., Ueno, H., Nakamura, Y., 1982. Synthesis and phase studies on sphalerite solid solution the systems Cu-Fe-Zn-S and Mn-Fe-Zn-S. Jpn. Assoc. Mineral. Petrol. Econ. Geol. 77, 271–280.

Shtrikman, H., Popovitz-Biro, R., Kretinin, A., Houben, L., Heiblum, M., Bukała, M., Galicka, M., Buczko, R., Kacman, P., 2009. Method for suppression of stacking faults in wurtzite III–V nanowires. Nano Lett. 9, 1506–1510.

Simpson, D.J., Bredow, T., Chandra, A.P., Cavallaro, G.P., Gerson, A.R., 2011. The effect of iron and copper impurities on the wettability of sphalerite (110) surface. J. Comput. Chem. 32, 2022–2030.

Smith, F.G., 1955. Structure of zinc sulphide minerals. Am. Miner. 40, 658–675.

Song, W., Soo, L.S., Savini, M., Popp, L., Colvin, V.L., Segatori, L., 2014. Ceria nanoparticles stabilized by organic surface coatings activate the lysosome-autophagy system and enhance autophagic clearance. ACS Nano 8, 10328–10342.

Sun, W., Corni, S., di Felice, R., 2013. Reactivity of the ZnS(10$\bar{1}$0) Surface to small organic ligands by Density Functional Theory. J. Phys. Chem. C 117, 16034–16041.

Sunagawa, I., 1984. Growth of crystals in nature. In: Part, I. (Ed.), Material Science of the Earth's Interior. Sunagawa I. Tokyo, Terrapub, pp. 63–105.

Tauson, V.L., 1978. Phase relations and structural characteristics of mixed crystals in the system ZnS-MnS. Geochem. Int. 1978, 33–45.

Tauson, V.L., Abramovich, M.G., 1988. Physicochemical transformations of real crystals in mineral systems. Izdvo "Nauka", Sibirskoe Otdnie, Novosibirsk, p. 121.

Tauson, V.L., Chernyshev, L.V., 1978. The sphalerite-wurtzite transition in isomorphous mixtures of the system ZnS-MnS-CdS-FeS. Geochem. Int. 15, 33–41.

Toby, B.H., von Dreele, R.B., 2013. GSAS-II: the genesis of a modern open-source all purpose crystallography software package. J. Appl. Crystallogr. 46, 544–549.

Ushakov, S.V., Navrotsky, A., 2005. Direct measurements of water adsorption enthalpy on hafnia and zirconia. Appl. Phys. Lett. 87, 164103.

Viets, J.G., Hopkins, R.T., Miller, B.M., 1992. Variations in minor and trace metals in sphalerite from mississippi valley-type deposits of the Ozark region; genetic implications. Econ. Geol. 87, 1897–1905.

Walker, B., Tamayo, A.B., Dang, X.-D., Zalar, P., Seo, J.H., Garcia, A., Tantiwiwat, M., Nguyen, T.-Q., 2009. Nanoscale phase separation and high photovoltaic efficiency in solution-processed, small-molecule bulk heterojunction solar cells. Adv. Funct. Mater. 19, 3063–3069.

Wright, K., Watson, G.W., Parker, S.C., Vaughan, D.J., 1998. Simulation of the structure and stability of sphalerite (ZnS) surfaces. Am. Miner. 83, 141–146.

Yoshiyama, H., Tanaka, S., Mikami, Y., Ohshio, S., Nishiura, J., Kawakami, H., Kobayashi, H., 1988. Role of surface energy in thin-film growth of electroluminescent ZnS, CaS and SrS. J. Cryst. Growth 86, 56–60.

Zhang, H., Banfield, J.F., 2004. Aggregation, coarsening, and phase transformation in ZnS nanoparticles studied by molecular dynamics simulations. Nano Lett. 4, 713–718.

Zhang, H., Huang, F., Gilbert, B., Banfield, J.F., 2003a. Molecular dynamics simulations, thermodynamic analysis, and experimental study of phase stability of zinc sulfide nanoparticles. J. Phys. Chem. B 107, 13051–13060.

Zhang, H., Gilbert, B., Huang, F., Banfield, J.F., 2003b. Water-driven structure transformation in nanoparticles at room temperature. Nature 424, 1025–1029.

Zhang, H., Chen, B., Gilbert, B., Banfield, J.F., 2006. Kinetically controlled formation of a novel nanoparticulate ZnS with mixed cubic and hexagonal stacking. J Mater. Chem. 16, 249–254.